\documentclass[preprint,secnumarabic,amssymb, nobibnotes, a4paper, aps, prb]{revtex4-1}

\usepackage[tbtags,sumlimits,nointlimits,reqno]{amsmath}
\usepackage{amssymb}
\usepackage{color}
\usepackage{float}
\usepackage{graphicx}
\usepackage{epstopdf}  
\usepackage{psfrag}
\usepackage{epsfig}
\usepackage{subfigure}
\usepackage{mathtools}
\usepackage{import}
\usepackage[utf8]{inputenc}

\newcommand{\env}[1]{\texttt{#1}}
\begin{document}

\title{Graphene Transverse Electric Surface Plasmon Detection\\using Nonreciprocity Modal Discrimination}%

\author{Nima Chamanara and Christophe Caloz}%
\affiliation{École Polytechnique de Montréal, Montréal, QC H3T 1J4, Canada.}
\date{July 10, 2016}%

\begin{abstract}
We present a magnetically biased graphene-ferrite structure discriminating the TE and TM plasmonic modes of graphene. In this structure, the graphene TM plasmons interact reciprocally with the structure. In contrast, the graphene TE plasmons exhibit nonreciprocity. This nonreciprocity is manifested in unidirectional TE propagation in a frequency band close to the interband threshold frequency. The proposed structure provides a unique platform for the experimental demonstration of the unusual existence of the TE plasmonic mode in graphene. 
\end{abstract}

\maketitle


\section{Introduction}
Graphene plasmonics has been an area of extensive research in the past few years \cite{grigorenko2012graphene,neto2009electronic,geim2007riseOfGraphene, politano2014plasmon, luo2013plasmons, chamanara2013_opex_coupler, chamanara2013_opex_isol}. Graphene plasmons have enabled photodetection enhancement \cite{mueller2010graphene}, light matter interaction enhancement in solar cells \cite{barnes2003surface}, and novel optical modulators and  sensors \cite{grigorenko2012graphene, liu2011graphene}. The Dirac band structure endows graphene with tunability, not easily obtainable in other plamonic materials \cite{liu2011graphene}. Moreover, the linearity of this band structure leads to the existence of unusual plasmonic modes that are unique to graphene \cite{mikhailov2007newMode}. It was shown theoretically in Ref.~\cite{mikhailov2007newMode, bordag2014transverse} that, in addition to the conventional transverse magnetic (TM) or longitudinal plasmonic mode, graphene also supports an unusual plasmonic mode which is transverse electric (TE). Compared to the conventional TM plasmons, the TE mode is more loosly confined to graphene, has a lower loss and propagates with a faster phase velocity \cite{hanson2008dyadic, kotov2013ultrahigh, he2014comparison, drosdoff2014transverse, werra2016te, chamanara2015fermat}. The TE mode can be excited when the imaginary part of the conductivity acquires a non-Drude sign. This condition is satisfied when the interband conductivity of graphene becomes dominant over its intraband conductivity, which is normally satisfied in a frequency window close to the interband transition threshold frequency \cite{mikhailov2007newMode, hanson2008dyadic}, that can be tuned from the microwave to the infrared frequency bands by adjusting graphene's chemical potential.

\begin{figure}[h!]
\psfrag{x}[c][c][0.9]{$x$}
\psfrag{y}[c][c][0.9]{$y$}
\psfrag{z}[c][c][0.9]{$z$}
\psfrag{a}[c][c][0.9]{TE magnetic field lines}
\psfrag{b}[l][c][0.9]{TE current density}
\psfrag{c}[c][b][0.9]{ferrite substrate}
\psfrag{d}[c][c][0.9]{magnetic bias}
\psfrag{e}[c][t][0.9]{ferrite superstrate}
\includegraphics[width=\columnwidth]{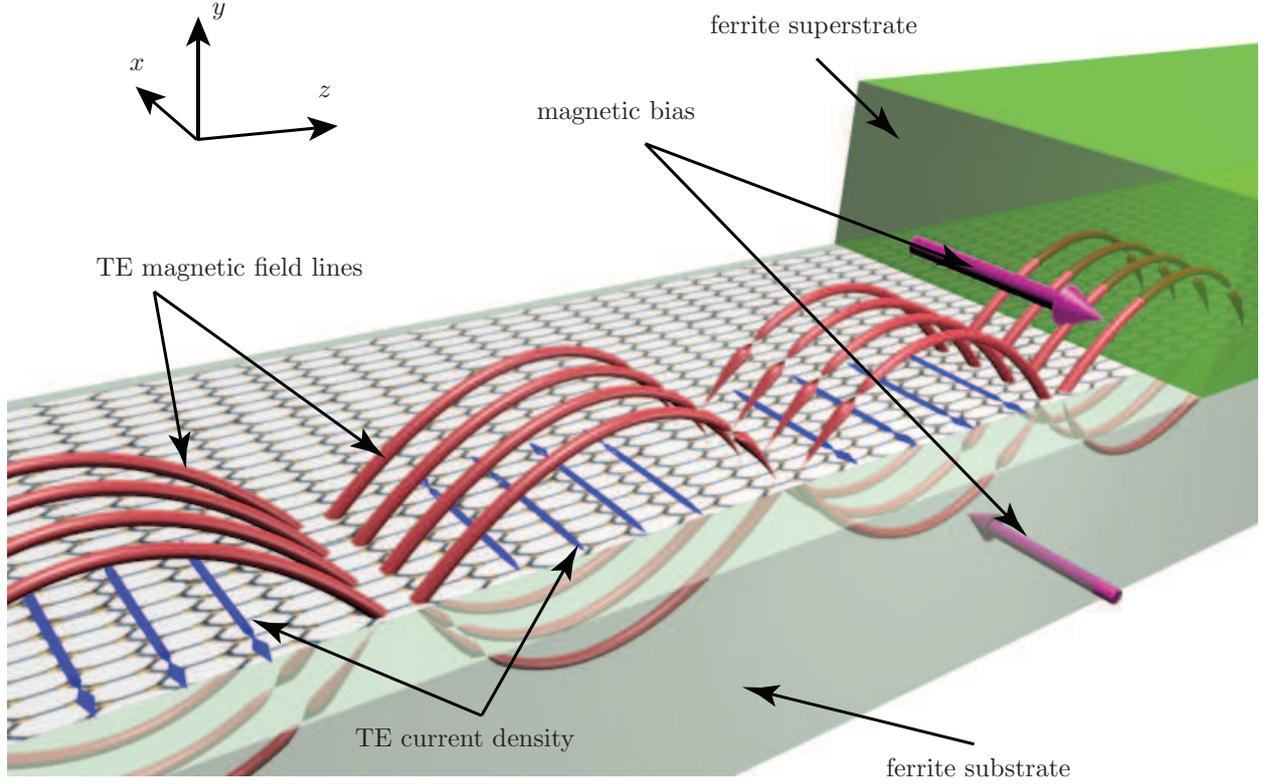}
\caption{Artistic representation of the TE surface plasmons in a graphene sheet sandwiched between a ferrite substrate and superstrate and propagating in the $+z$ direction. The electric current is transverse and sinusoidally varying along graphene, with the magnetic fields looping around them. The electric field (not shown) is completely transverse. The magnetic bias field is parallel to the plane of graphene and looking in opposite directions in the substrate a superstrate. For clarity only part of the superstrate is shown.}
\label{fig:graph_ferrite_persp}
\end{figure}

The specific field configuration of graphene TE plasmons leads to nonreciprocal interaction with a magnetically biased ferrite substrate or superstrate, whereas graphene TM plasmons do not nonreciprocally interact with such a structure. The magnetic field lines of the graphene TE mode and its electric current are shown in Fig.~\ref{fig:graph_ferrite_persp}, for propagation along the $z$ direction. The electric current is transverse to the direction of propagation, with sinusoidal variation along~$z$. The magnetic field lines loop around current sections, as shown in Fig.~\ref{fig:graph_ferrite_persp}, and the electric field (not shown in the figure) is completely transverse. Such a magnetic field generally interact nonreciprocally with a properly magnetized ferrite substrate/superstrate. In order to include nonreciprocal interaction, the ferrite substrate/superstrate should be biased by a static magnetic field parallel to the plane of graphene and normal to the direction of propagation, as shown in Fig.~\ref{fig:graph_ferrite_persp}. The magnetically biased ferrite substrate/superstrate acquires then a tensorial permeability in the $yz$ plane. However, the conductivity of graphene is scalar (no cyclotron orbiting) since the magnetic field is parallel to its plane.

\begin{figure}[h!]
\psfrag{y}[c][c][0.9]{$y$}
\psfrag{z}[c][c][0.9]{$z$}
\psfrag{a}[c][b][0.9]{$A$}
\psfrag{c}[c][c][0.9]{$A$}
\psfrag{b}[c][t][0.9]{$B$}
\psfrag{d}[c][c][0.9]{$B$}
\psfrag{f}[c][c][0.8]{forward $+z$}
\psfrag{e}[c][c][0.8]{backward $-z$}
\psfrag{g}[c][c][1.0]{$\epsilon_{r1}, \bar{\bar\mu}_{r1}$}
\psfrag{h}[c][c][1.0]{$\epsilon_{r2}, \bar{\bar\mu}_{r2}$}
\includegraphics[width=0.95\columnwidth]{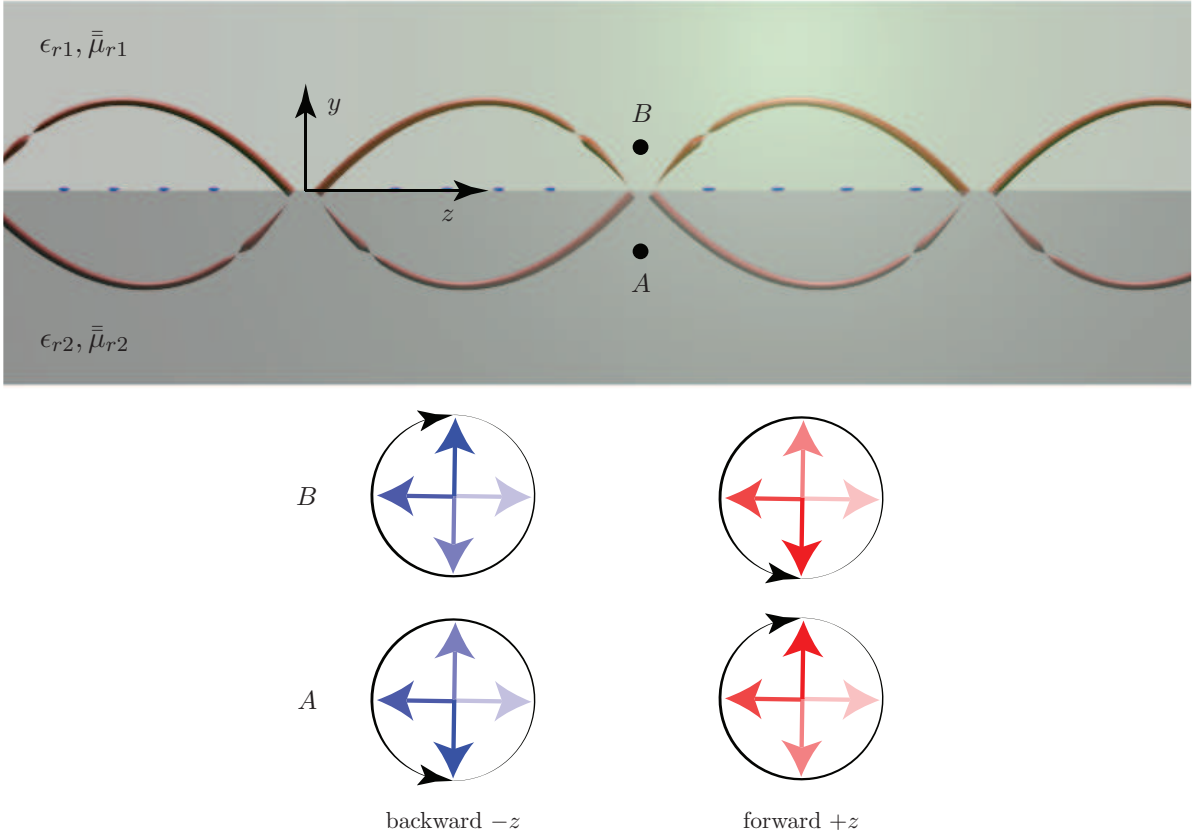}
\caption{Longitudinal ($yz$) cross-section of the structure in Fig.~\ref{fig:graph_ferrite_persp}. Any point A/B perceives clockwise/counterclockwise and counterclockwise/clockwise rotating magnetic fields, respectively, as the wave propagates in the $+z$ direction and in the the $-z$ direction, respectively.}
\label{fig:graph_ferrite_side}
\end{figure}

Figure.~\ref{fig:graph_ferrite_side} shows a longitudinal ($yz$) cross-section of the structure in Fig.~\ref{fig:graph_ferrite_persp}, with the arrows representing the magnetic field lines. As the magnetic field propagates along graphene, any point $A$/$B$ inside the ferrite substrate/superstrate sees a rotating magnetic field in the $yz$ plane, with clockwise/counterclockwise rotation for forward and counterclockwise/clockwise rotation for the backward direction. For an antisymmetric permeability tensor $\mu = \mu_d(\mathbf{y}\mathbf{y}+\mathbf{z}\mathbf{z}) + \mu_o(\mathbf{y}\mathbf{z}-\mathbf{z}\mathbf{y})$, such left and right handed rotating magnetic fields perceive effective scalar permeabilities $\mu_d+i\mu_o$ and $\mu_d-i\mu_o$, respectively. For the magnetic bias configuration shown in Fig.~\ref{fig:graph_ferrite_persp}, the substrate and superstrate exhibit opposite sign off-diagonal permeability components \mbox{($\mu_{o}^A=-\mu_{o}^B$)}. For a TE wave propagating in the forward direction point A (right handed) and B (left handed) perceive \mbox{$\mu^A = \mu^A_d+i\mu^A_o$} and \mbox{$\mu^B = \mu^B_d-i\mu^B_o$} effective permeabilities, respectively. Since \mbox{$\mu_{o}^A=-\mu_{o}^B$} the effective permeability perceived by both points is \mbox{$\mu^A_d+i\mu^A_o$}. Similarly for the backward direction the effective permeability perceived by both points is \mbox{$\mu^A_d-i\mu^A_o$}. Therefore, the ferrite structure is effectively seen as different media for opposite direction of propagation \cite{lax1962microwave}, and therefore exhibits nonreciprocity.

\section{Analysis}

For an infinite graphene sheet between two semi-infinite ferrite media, the electromagnetic fields supported by the structure and their dispersion relations can be derived analytically. The plasmonic electric fields in regions 1 (above graphene) and 2 (below graphene), are expressed as surface waves propagating along the $z$ direction with propagation constant $k$, and exponentially decaying in the $+y$ and $-y$ directions with decay rates $\alpha_1$ and $\alpha_2$, respectively ($e^{+i\omega t}$ convention):

\begin{subequations} \label{eq:electric_field}
\begin{equation}
\mathbf{\mathbf{E}_{1}}=\left[\begin{matrix}E_{{x1}}, & E_{{y1}}, & E_{{z1}}\end{matrix}\right]e^{-\alpha_{1}y-ikz},
\end{equation}
\begin{equation}
\mathbf{E}_{\mathbf{2}}=\left[\begin{matrix}E_{{x2}}, & E_{{y2}}, & E_{{z2}}\end{matrix}\right]e^{\alpha_{2}y-ikz},
\end{equation}
\end{subequations}

\noindent
For generality of the analysis it is initially assumed that the direction of the magnetic bias in regions 1 and 2 are not necessarily opposite and may take arbitrary values. With the relative permeability tensors
 
\begin{equation} \label{eq:permeability_tensor}
\bar{\bar{\mu}}_{r1} = \left[\begin{matrix}1 & 0 & 0\\0 & \mu_{{rd1}} & - \mu_{{ro1}}\\0 & \mu_{{ro1}} & \mu_{{rd1}}\end{matrix}\right],~~~
\bar{\bar{\mu}}_{r2} = \left[\begin{matrix}1 & 0 & 0\\0 & \mu_{{rd2}} & - \mu_{{ro2}}\\0 & \mu_{{ro2}} & \mu_{{rd2}}\end{matrix}\right]
\end{equation}

\noindent
of the $x$-biased superstrate and substrate ferrite material, the magnetic fields above and below graphene are found through Maxwell equations as

\begin{equation} \label{eq:magnetic_field}
\mathbf{H}_{i}=\frac{-1}{i\omega\mu_{0}}\mathbf{\bar{\bar{\mu}}}_{ri}^{-1}\centerdot\nabla\times\mathbf{E}_{i}.
\end{equation}

\noindent
where the subscript $i=1,2$ represents the fields in regions 1 or 2. The eigenmodes and their dispersion are then found by the application of the boundary condition in the graphene plane,

\begin{equation} \label{eq:boundary_cond_magnetic}
\left.\mathbf{a}_{y}\times\left(\mathbf{H_{1}}-\mathbf{H_{2}}\right)\right|_{y=0}=\left.\mathbf{\sigma}\mathbf{E_{T}}\right|_{y=0},
\end{equation}

\noindent
where, by continuity of the electric field \mbox{$E_{x1}=E_{x2}$}, \mbox{$E_{z1}=E_{z2}$},  so that the tangential electric field simply reads \mbox{$\mathbf{E_T}=E_{x1}\mathbf{x}+E_{z1}\mathbf{z}$}. Substituting \eqref{eq:electric_field} and \eqref{eq:magnetic_field} in \eqref{eq:boundary_cond_magnetic} results in

\begin{subequations} \label{eq:matrix_Exyz}
\begin{equation}
\begin{split}
E_{{x1}}\mu_{0}\omega\sigma\left(\mu_{{rd1}}^{2}+\mu_{{ro1}}^{2}\right)\left(\mu_{{rd2}}^{2}+\mu_{{ro2}}^{2}\right)+\\
E_{{x1}}\left(\mu_{{rd2}}^{2}+\mu_{{ro2}}^{2}\right)\left(i\alpha_{1}\mu_{{rd1}}-\mu_{{ro1}}k\right)+\\
E_{{x2}}\left(\mu_{{rd1}}^{2}+\mu_{{ro1}}^{2}\right)\left(i\alpha_{2}\mu_{{rd2}}+\mu_{{ro2}}k\right)=0
\end{split},
\end{equation}
\begin{equation}
E_{{y1}}k-E_{{y2}}k+iE_{{z1}}\alpha_{1}-E_{{z1}}\mu_{0}\omega\sigma+iE_{{z2}}\alpha_{2}=0.
\end{equation}
\end{subequations}

\noindent
The normal electric field components $E_{y1}$ and $E_{y2}$ are redundant, as they may be expressed in terms of $E_{z1}$ and $E_{z2}$, respectively, through the divergence relations $\nabla\centerdot\mathbf{E_1} = 0$ and $\nabla\centerdot\mathbf{E_2} = 0$, as

\begin{equation} \label{eq:Ey}
E_{y1} =- \frac{i E_{{z1}}}{\alpha_{1}} k,~~~~ E_{y2} =\frac{i E_{{z2}}}{\alpha_{2}} k,
\end{equation}

\noindent
leading to 

\begin{subequations} \label{eq:matrix_Exz}
\begin{equation} 
\begin{split}
\left[-\mu_{0}\omega\sigma\left(\mu_{{rd1}}^{2}+\mu_{{ro1}}^{2}\right)\left(\mu_{{rd2}}^{2}+\mu_{{ro2}}^{2}\right)+\right.&\\
\left(\mu_{{rd1}}^{2}+\mu_{{ro1}}^{2}\right)\left(i\alpha_{2}\mu_{{rd2}}+\mu_{{ro2}}k\right)+&\\
\left.\left(\mu_{{rd2}}^{2}+\mu_{{ro2}}^{2}\right)\left(i\alpha_{1}\mu_{{rd1}}-\mu_{{ro1}}k\right)\right] E_{x1}=0&,
\end{split}
\end{equation}
\begin{equation}
\left(\frac{i\alpha_{1}}{\mu_{0}\omega}+\frac{i\alpha_{2}}{\mu_{0}\omega}-\sigma-\frac{ik^{2}}{\alpha_{2}\mu_{0}\omega}-\frac{ik^{2}}{\alpha_{1}\mu_{0}\omega}\right)E_{z1}=0.
\end{equation}
\end{subequations}

\noindent
The decay rates $\alpha_1$ and $\alpha_2$ may be expressed in terms of the propagation constant $k$, through the electric field wave equation in regions 1 and 2,

\begin{subequations} \label{eq:wave_eq}
\begin{equation} \label{eq:wave_eq_1}
\nabla\times\mathbf{\bar{\bar{\mu}}}_{r1}^{-1}\centerdot\nabla\times\mathbf{E_{1}}-
\omega^{2}\mu_{0}\varepsilon_{0}\varepsilon_{r1}\mathbf{E_{1}}=\mathbf{0},
\end{equation}
\begin{equation} \label{eq:wave_eq_2}
\nabla\times\mathbf{\bar{\bar{\mu}}}_{r2}^{-1}\centerdot\nabla\times\mathbf{E_{2}}-
\omega^{2}\mu_{0}\varepsilon_{0}\varepsilon_{r2}\mathbf{E_{2}}=\mathbf{0},
\end{equation}
\end{subequations}

\noindent
which enforce the relations

\begin{widetext}
\begin{equation} \label{eq:wave_eq_expanded_1}
\left[\begin{matrix}\frac{E_{{x1}}}{\mu_{{rd1}}^{2}+\mu_{{ro1}}^{2}}\left(-\alpha_{1}^{2}\mu_{{rd1}}-\epsilon_{0}\epsilon_{{r1}}\mu_{0}\mu_{{rd1}}^{2}\omega^{2}-\epsilon_{0}\epsilon_{{r1}}\mu_{0}\mu_{{ro1}}^{2}\omega^{2}+\mu_{{rd1}}k^{2}\right)\\
\frac{iE_{{z1}}}{\alpha_{1}}k\left(\alpha_{1}^{2}+\epsilon_{0}\epsilon_{{r1}}\mu_{0}\omega^{2}-k^{2}\right)\\
E_{{z1}}\left(-\alpha_{1}^{2}-\epsilon_{0}\epsilon_{{r1}}\mu_{0}\omega^{2}+k^{2}\right)
\end{matrix}\right]=\mathbf{0},
\end{equation}

\begin{equation} \label{eq:wave_eq_expanded_2}
\left[\begin{matrix}\frac{E_{{x2}}}{\mu_{{rd2}}^{2} + \mu_{{ro2}}^{2}} \left(- \alpha_{2}^{2} \mu_{{rd2}} - \epsilon_{0} \epsilon_{{r2}} \mu_{0} \mu_{{rd2}}^{2} \omega^{2} - \epsilon_{0} \epsilon_{{r2}} \mu_{0} \mu_{{ro2}}^{2} \omega^{2} + \mu_{{rd2}} k^{2}\right)\\\frac{i E_{{z2}}}{\alpha_{2}} k \left(- \alpha_{2}^{2} - \epsilon_{0} \epsilon_{{r2}} \mu_{0} \omega^{2} + k^{2}\right)\\E_{{z2}} \left(- \alpha_{2}^{2} - \epsilon_{0} \epsilon_{{r2}} \mu_{0} \omega^{2} + k^{2}\right)\end{matrix}\right]=\mathbf{0},
\end{equation}
\end{widetext}

\noindent
where $\epsilon_{r1}$ and $\mathbf{\bar{\bar{\mu}}}_{r1}$ are the relative permittivity and the relative permeability tensor in region 1, respectively, and $\epsilon_{r2}$ and $\mathbf{\bar{\bar{\mu}}}_{r2}$ are the relative permittivity and the relative permeability tensor in region 2, respectively.

Equations \eqref{eq:matrix_Exz}, \eqref{eq:wave_eq_expanded_1} and \eqref{eq:wave_eq_expanded_2} describe the eigenmodes of the system. This set of equations admits two modal solutions. The first solution is a TM mode, for which $E_x=0$ and

\begin{subequations}
\begin{equation}
\alpha_1 = \sqrt{- \epsilon_{0} \epsilon_{{r1}} \mu_{0} \omega^{2} + k^{2}},
\end{equation}
\begin{equation}
\alpha_2 = \sqrt{- \epsilon_{0} \epsilon_{{r2}} \mu_{0} \omega^{2} + k^{2}}.
\end{equation}
\end{subequations}

\noindent
This mode has its magnetic field along the DC magnetic bias, and therefore sees the ferrite as an isotropic medium with scalar permeability $\mu_0$. It therefore interacts reciprocally with the ferrite, with identical characteristics in the forward and backward directions. The dispersion relation for this mode is given by

\begin{equation} \label{eq:TM_dispersion}
\sigma+\frac{i\epsilon_{0}\epsilon_{{r1}}\omega}{\sqrt{-\epsilon_{0}\epsilon_{{r1}}\mu_{0}\omega^{2}+k^{2}}}+\frac{i\epsilon_{0}\epsilon_{{r2}}\omega}{\sqrt{-\epsilon_{0}\epsilon_{{r2}}\mu_{0}\omega^{2}+k^{2}}}=0.
\end{equation}

\noindent
The off-diagonal component of the permeability tensor has no contribution, confirming reciprocal interaction with the magnetically biased ferrite structure. The dispersion is a function of $k^2$ and is therefore reciprocal with respect to the direction of propagation, as expected. 

The second solution is a TE surface plasmon mode ($E_z=0$), with decay rates in the normal direction

\begin{subequations}
\begin{equation}
\alpha_1 = \sqrt{- \epsilon_{0} \epsilon_{{r1}} \mu_{0} \mu_{{rd1}} \omega^{2} - \frac{\epsilon_{0} \epsilon_{{r1}}}{\mu_{{rd1}}} \mu_{0} \mu_{{ro1}}^{2} \omega^{2} + k^{2}},
\end{equation}
\begin{equation}
\alpha_2 = \sqrt{- \epsilon_{0} \epsilon_{{r2}} \mu_{0} \mu_{{rd2}} \omega^{2} - \frac{\epsilon_{0} \epsilon_{{r2}}}{\mu_{{rd2}}} \mu_{0} \mu_{{ro2}}^{2} \omega^{2} + k^{2}}.
\end{equation}
\end{subequations}

\noindent
This mode has its magnetic field perpendicular to the DC magnetic bias. As explained above, such a magnetic field perceives different effective materials in the forward and backward directions and is thus nonreciprocal. The dispersion relation for the TE mode is

\begin{widetext}
\begin{equation} \label{eq:TE_dispersion}
\begin{split}
&-\mu_{0}\omega\sigma\left(\mu_{{rd1}}^{2}+\mu_{{ro1}}^{2}\right)\left(\mu_{{rd2}}^{2}+\mu_{{ro2}}^{2}\right)+\\
&\left(\mu_{{rd1}}^{2}+\mu_{{ro1}}^{2}\right)\left(i\mu_{{rd2}}\sqrt{\frac{1}{\mu_{{rd2}}}\left(-\epsilon_{0}\epsilon_{{r2}}\mu_{0}\mu_{{ro2}}^{2}\omega^{2}+\mu_{{rd2}}\left(-\epsilon_{0}\epsilon_{{r2}}\mu_{0}\mu_{{rd2}}\omega^{2}+k^{2}\right)\right)}+\mu_{{ro2}}k\right)+\\
&\left(\mu_{{rd2}}^{2}+\mu_{{ro2}}^{2}\right)\left(i\mu_{{rd1}}\sqrt{\frac{1}{\mu_{{rd1}}}\left(-\epsilon_{0}\epsilon_{{r1}}\mu_{0}\mu_{{ro1}}^{2}\omega^{2}+\mu_{{rd1}}\left(-\epsilon_{0}\epsilon_{{r1}}\mu_{0}\mu_{{rd1}}\omega^{2}+k^{2}\right)\right)}-\mu_{{ro1}}k\right)=0.
\end{split}
\end{equation}
\end{widetext}

\noindent 
The term $\mu_{{ro2}} k$ in this relation, which is odd in $k$, results in a dispersion that is different for positive and negative $k$'s, corresponding to nonreciprocity. Note that if the substrate and superstrate ferrites are identical and have the same parallel magnetic bias, this dispersion relation remains symmetric with respect to propagation direction (opposite signs of $k$) and is thus reciprocal. In this case the nonreciprocity produced by the substrate and superstrate cancel out each other. To generate nonreciprocity the magnetic bias should be different. Oppositely directed magnetic fields generate maximum nonreciprocity. This nonreciprocity is manifested in unidirectional TE plasmon propagation on a frequency band close to the intrband frequency threshold ($\omega=2\mu_c$). Although the magnetic effect produced by the ferrite is relatively weak at infrared and optical frequencies, we next show that over a specific frequency band which is tunable by graphene and ferrite parameters, isolation is theoretically infinite. Equation \eqref{eq:TE_dispersion} does not admit analytic solutions and should be solved numerically. Some guidelines regarding numerical solution of dispersion equations are provided in the supplementary material~\cite{chamanaraSupplementalMat}.

\begin{figure}[ht!]
\subfigure[]{\label{fig:cond_intra_inter}
\includegraphics[width=1.0\columnwidth]{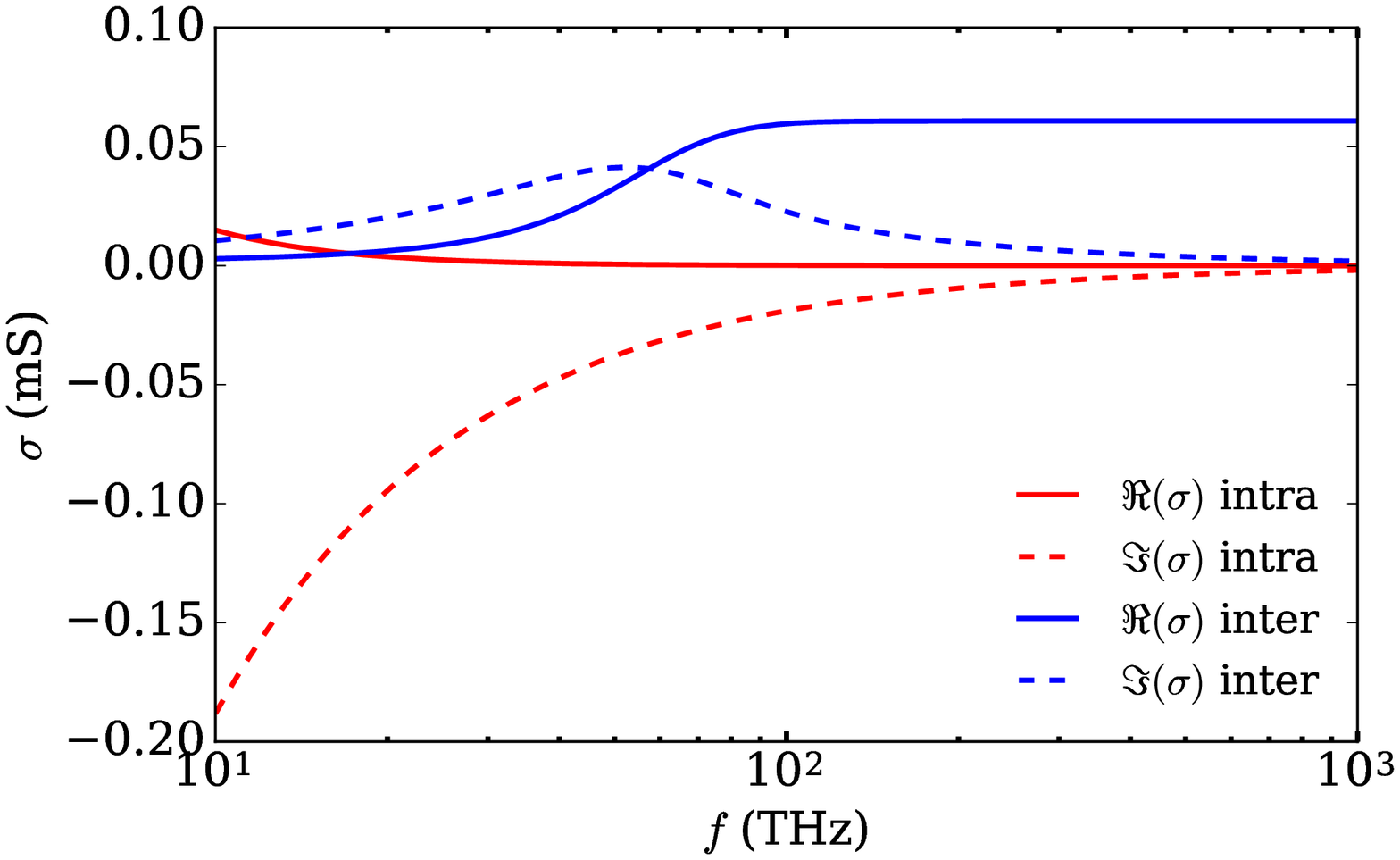}}
\subfigure[]{\label{fig:cond_total}
\includegraphics[width=1.0\columnwidth]{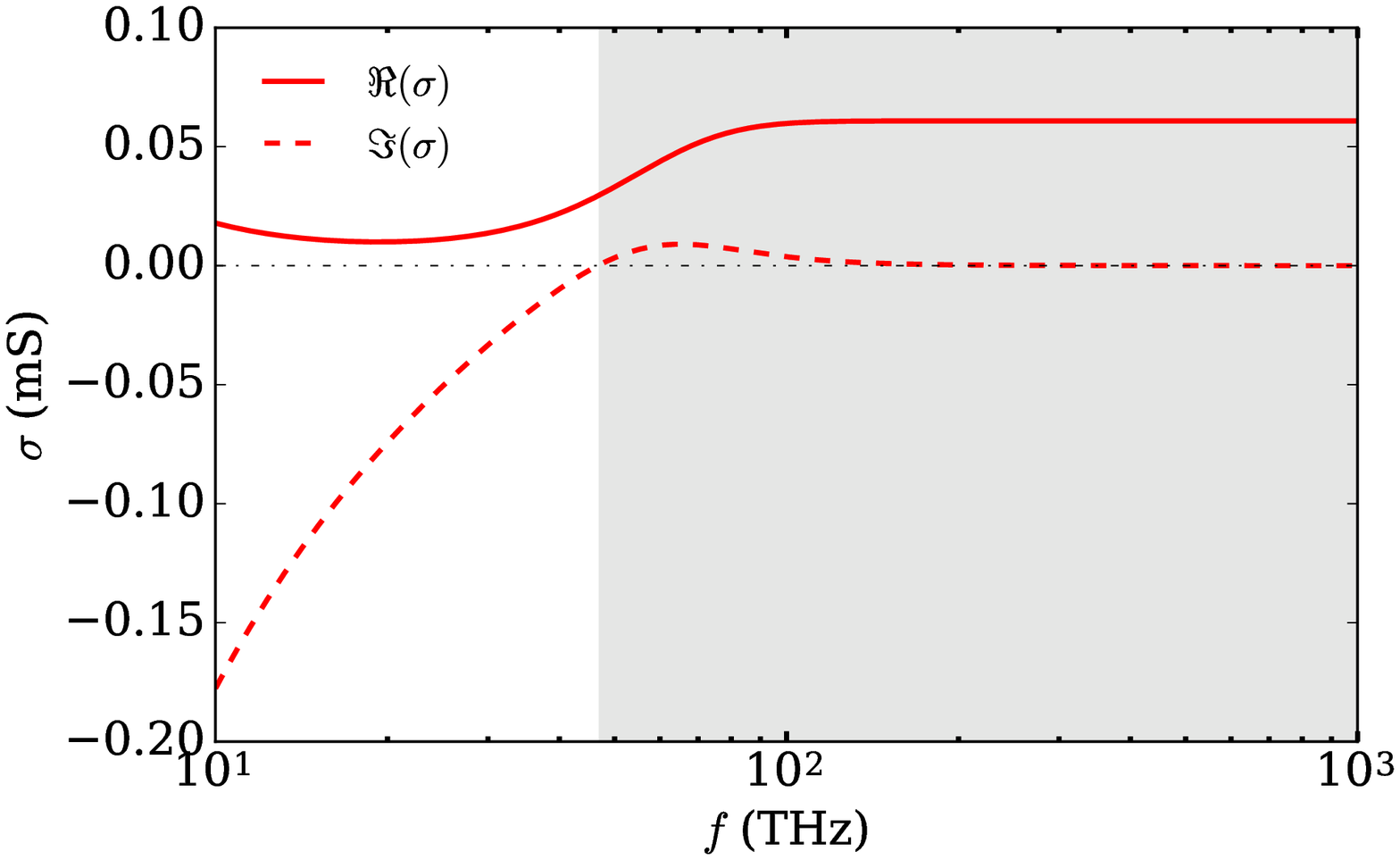}}
\caption{Conductivity of a graphene sheet with chemical potential $\mu_c=0.1$~eV, scattering time $\tau=0.2$~ps, and temperature $T=300$~K. (a) Intraband and interband conductivities plotted separately. (b) Total conductivity. The interband conductivity becomes dominant in the highlighted frequency range where the imaginary part of the total conductivity flips sign.}
\label{fig:conductivity}
\end{figure}

\section{Results}

Consider a graphene sheet with chemical potential $\mu_c=0.1$~eV, scattering time $\tau=0.2$~ps, and temperature $T=300$~K. The corresponding interband and intraband conductivities \cite{gusynin2007magneto, gusynin2006transport, gusynin2009universal} are plotted in Fig.~\ref{fig:cond_intra_inter}. Close to the interband frequency threshold ($\omega=2\mu_c$), the interband conductivity becomes dominant over the intraband conductivity, and the imaginary part of the total conductivity flips sign, as shown in Fig.~\ref{fig:cond_total}. This region corresponds to the frequency band where the TE plasmonic mode of graphene can propagate along graphene.

\begin{figure}[ht!]\label{fig:TE-disp-sens}
\subfigure[]{\label{fig:TE-disp-freq}
\includegraphics[width=0.95\columnwidth]{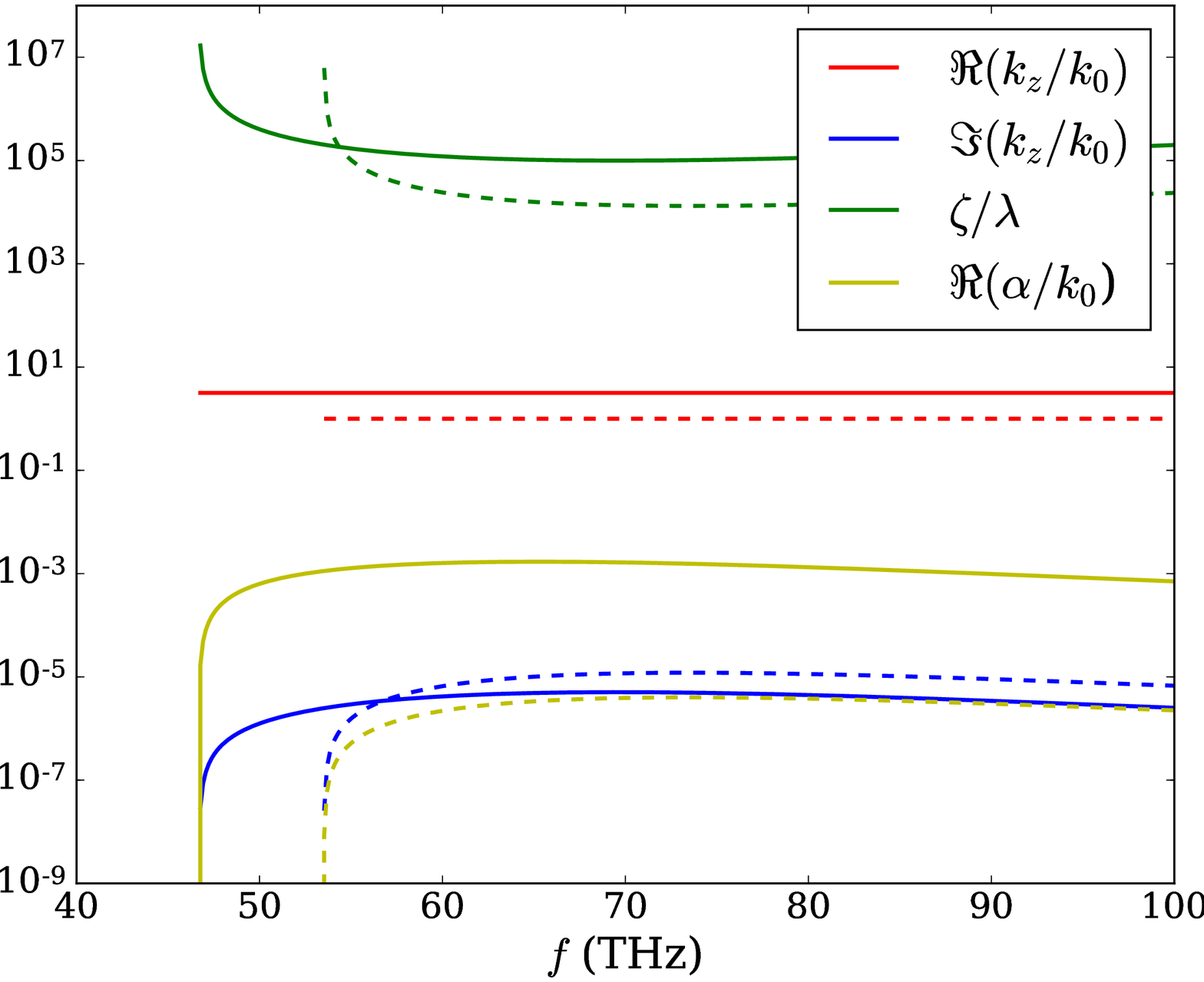}}
\subfigure[]{\label{fig:TE-depsdmu-region}
\includegraphics[width=1.0\columnwidth]{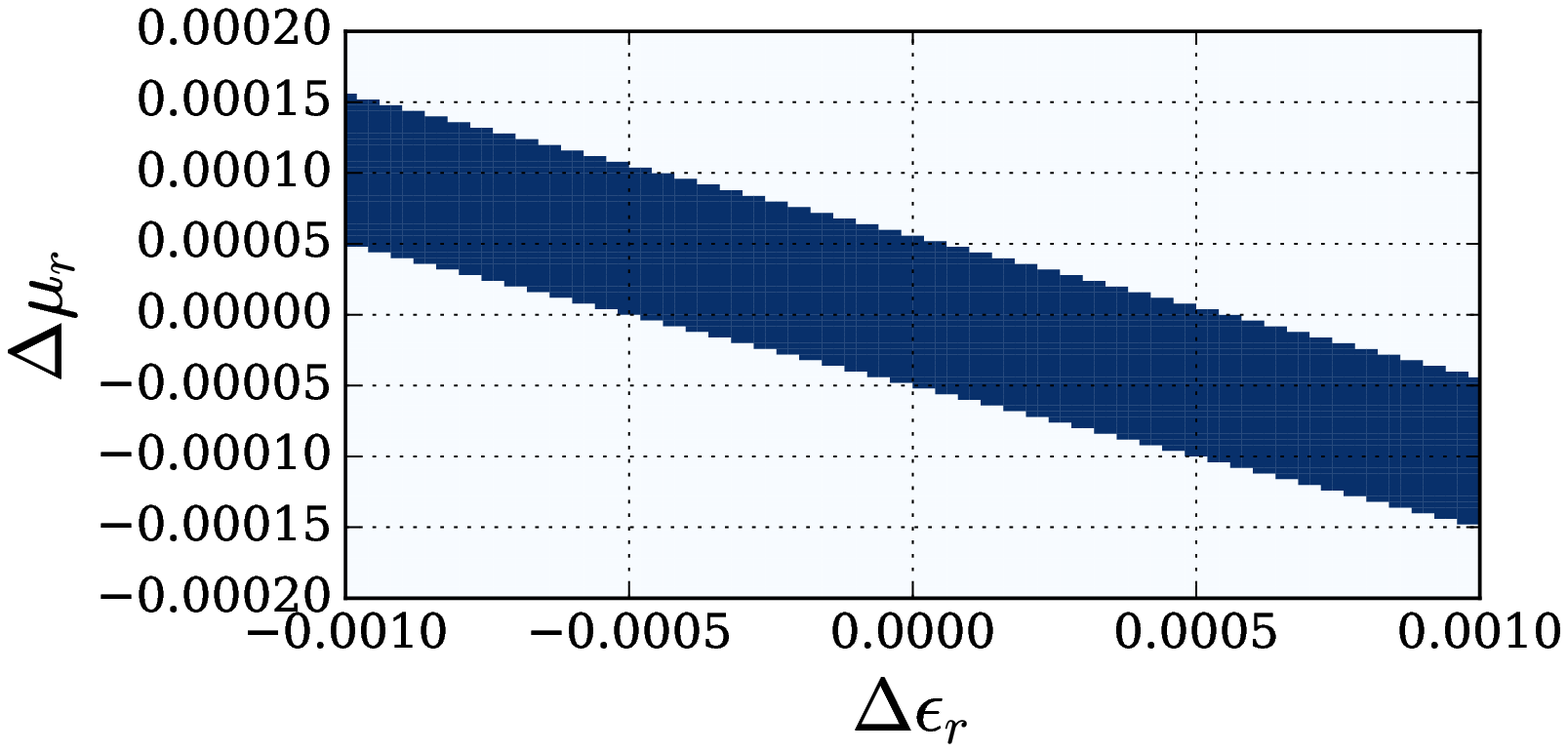}}
\caption{ (a) TE normalized propagation constant, loss, propagation length and confinement factor for a graphene sheet with chemical potential $\mu_c=0.1$~eV, scattering time $\tau=0.2$~ps, and temperature $T=300$~K. The solid lines represent the dispersion for $\epsilon_{r1}=10.0$, $\mu_{r1}=1$, $\epsilon_{r2}=10.0$, $\mu_{r2}=1$ and dashed lines for $\epsilon_{r1}=10.0$, $\mu_{r1}=1$, $\epsilon_{r2}=10.0001$, $\mu_{r2}=1.00001$. (b) Sensitivity of TE surface plasmons with respect to material contrast. The TE mode is supported only for material contrasts inside the highlighted strip.}
\end{figure}

First consider a graphene sheet sandwiched between two magnetically unbiased media with scalar parameters $\epsilon_{r1}, \mu_{r1}$ and $\epsilon_{r2}, \mu_{r2}$. The TE dispersion equations for such a structure is 

\begin{equation} \label{eq:TE_disp_scalar}
k=\sqrt{\epsilon_{r1}\mu_{r1}k_{0}^{2}+\alpha_{1}^{2}}=\sqrt{\epsilon_{r2}\mu_{r2}k_{0}^{2}+\alpha_{2}^{2}},
\end{equation}

\noindent
where $k_0$ is the free space wave number and $\alpha_1$ and $\alpha_2$ are decay rates normal to graphene surface in regions 1 and 2 with the following relations (details provided in the supplementary material~\cite{chamanaraSupplementalMat})

\begin{subequations} \label{eq:TE_alpha_scalar}
\begin{equation}
\begin{split}
&\alpha_1/k_0 = \frac{\mu_{1}}{\mu_{2} \left(\mu_{1}^{2} - \mu_{2}^{2}\right)} \left(i \mu_{2}^{3} \sigma\eta_0 \pm \right.\\ 
&\left.\sqrt{\mu_{2}^{2} \left(- \epsilon_{1} \mu_{1}^{3} + \epsilon_{1} \mu_{1} \mu_{2}^{2} + \epsilon_{2} \mu_{1}^{2} \mu_{2} - \epsilon_{2} \mu_{2}^{3} - \mu_{1}^{2} \mu_{2}^{2} \sigma^{2}\eta_0^2\right)}\right),
\end{split}
\end{equation}
\begin{equation}
\begin{split}
&\alpha_2/k_0 = - \frac{1}{\mu_{1}^{2} - \mu_{2}^{2}} \left(i \mu_{1}^{2} \mu_{2} \sigma\eta_0 \pm \right.\\
&\left.\sqrt{- \mu_{2}^{2} \left(\epsilon_{1} \mu_{1}^{3} - \epsilon_{1} \mu_{1} \mu_{2}^{2} - \epsilon_{2} \mu_{1}^{2} \mu_{2} + \epsilon_{2} \mu_{2}^{3} + \mu_{1}^{2} \mu_{2}^{2} \sigma^{2}\eta_0^2\right)}\right).
\end{split}
\end{equation}
\end{subequations}

\noindent
where $\mu_i=\mu_{ri}$, $\epsilon_i=\epsilon_{ri}$ and $\eta_0=\sqrt{\mu_0/\epsilon_0}$. For $\epsilon_{r1}=\epsilon_{r2}$ and $\mu_{r1}=\mu_{r2}$ the solution to dispersion equations \eqref{eq:TE_disp_scalar}-\eqref{eq:TE_alpha_scalar} for $\Im{\sigma}>0$ lies in the proper Riemann sheet ($\Re{\alpha_i}>0$), corresponding to a surface wave with exponential decay normal to the graphene sheet. Therefore for identical media on both sides of graphene, the TE mode propagates at all the frequencies marked by grey color in Fig.~\ref{fig:cond_total}. For $\epsilon_{r1}=\epsilon_{r2}=10$ and and $\mu_{r1}=\mu_{r2}=1$ the corresponding TE normalized phase constant $\Re{k_z}/k_0$, loss $\Im{k_z}/k_0$, propagation length $\zeta = -1/\Im{k_z}$ and confinement factor $\alpha/k_0$ are presented in Fig.~\ref{fig:TE-disp-freq} by solid curves. The dashed curves correspond to a very small change in the permittivity and permeability of region 2 by $10^{-3}$ and $10^{-4}$ respectively. The dispersion curves change dramatically for such a small contrast in material parameters. Therefore the TE mode is highly sensitive to the the contrast of the material parameters on both sides of graphene \cite{kotov2013ultrahigh}. As the contrast between the material parameters of regions 1 and 2 is increased, the solutions to dispersion equations \eqref{eq:TE_disp_scalar}-\eqref{eq:TE_alpha_scalar} quickly moves to the improper Riemann sheet and becomes unphysical. For  $\epsilon_{r1}=10$ and $\mu_{r1}=1$ the material contrast corresponding to proper surface wave solutions is plotted in Fig.~\ref{fig:TE-depsdmu-region}. For the structure of Fig.~\ref{fig:graph_ferrite_persp} the substrate and superstrate ferrites should be almost identical, otherwise the TE surface plasmon mode can not propagate.

\begin{figure}[ht!]
\includegraphics[width=1.0\columnwidth]{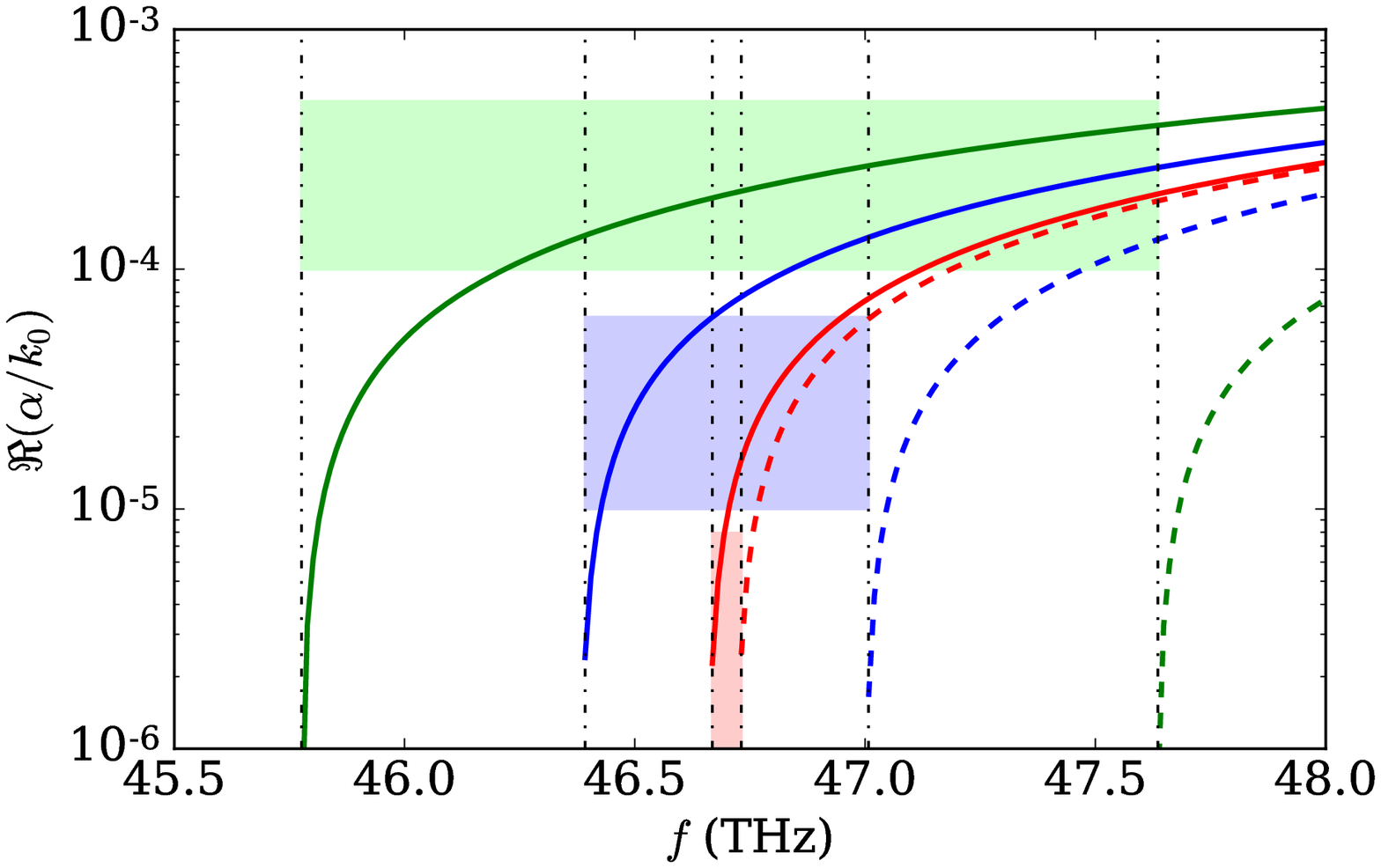}
\caption{Decay rate normal to graphene ($y$) for the forward (solid curved) and backward (dashed curves) TE surface plasmons, for a graphene sheet with chemical potential $\mu_c=0.1$~eV, scattering time $\tau=0.2$~ps, and temperature $T=300$~K, between two ferrite media with anti-parallel magnetic bias as in Fig.~\ref{fig:graph_ferrite_persp} and with ferromagnetic resonances $f_0=0.1$~GHz (red), $f_0=1.0$~GHz (blue) and $f_0=3.0$~GHz (green). The corresponding magnetic bias field is $B_0=0.00356$~T (red), $B_0=0.0356$~T (blue), $B_0=0.106$~T (green). A loss factor $\alpha=0.05$ is assumed for the ferrites.}
\label{fig:alpha_k0}
\end{figure}

\begin{figure}[]
\subfigure[]{\label{fig:slow_wave_factor}
\includegraphics[width=1.0\columnwidth]{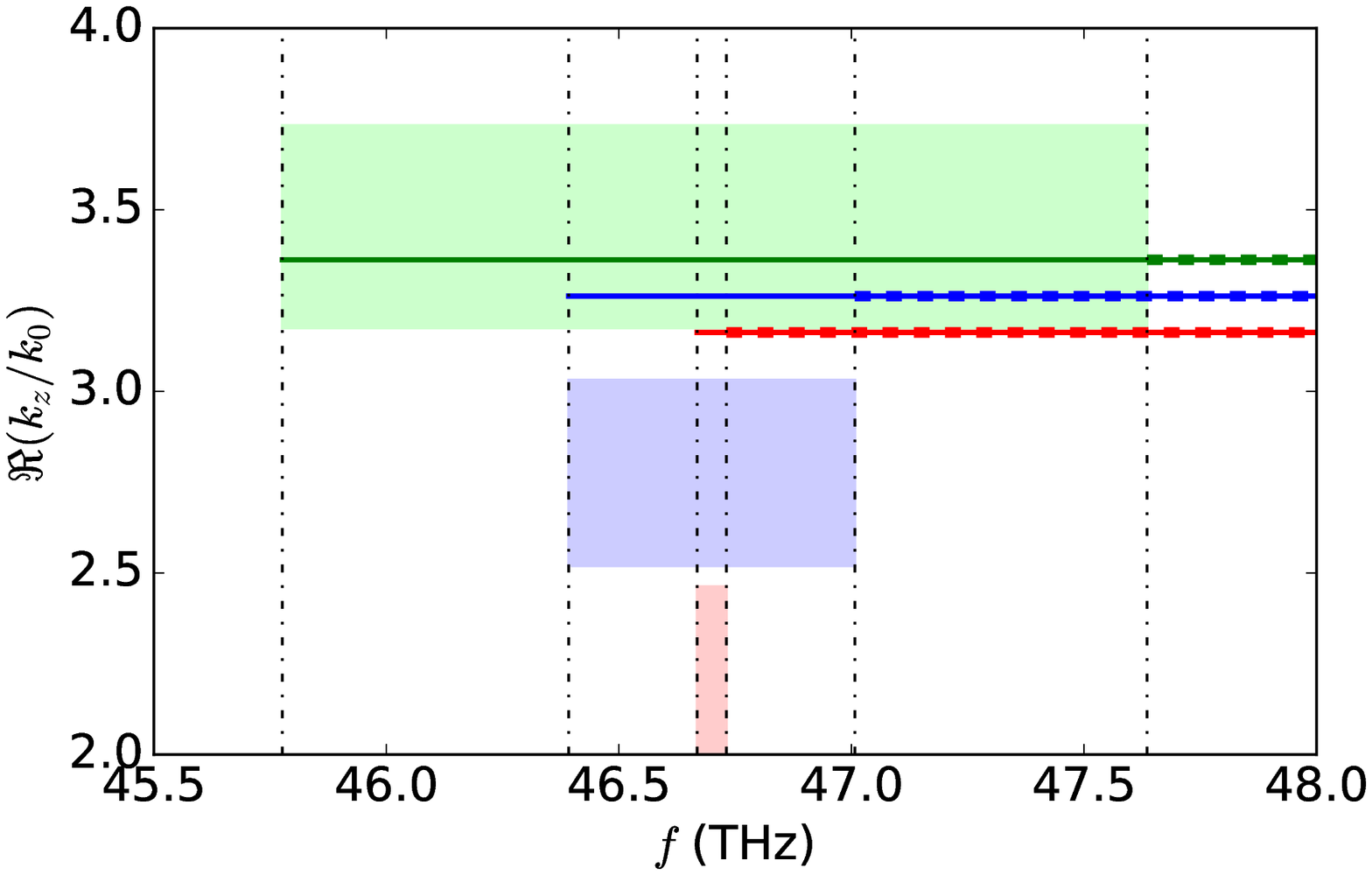}}
\subfigure[]{\label{fig:prop_length}
\includegraphics[width=1.0\columnwidth]{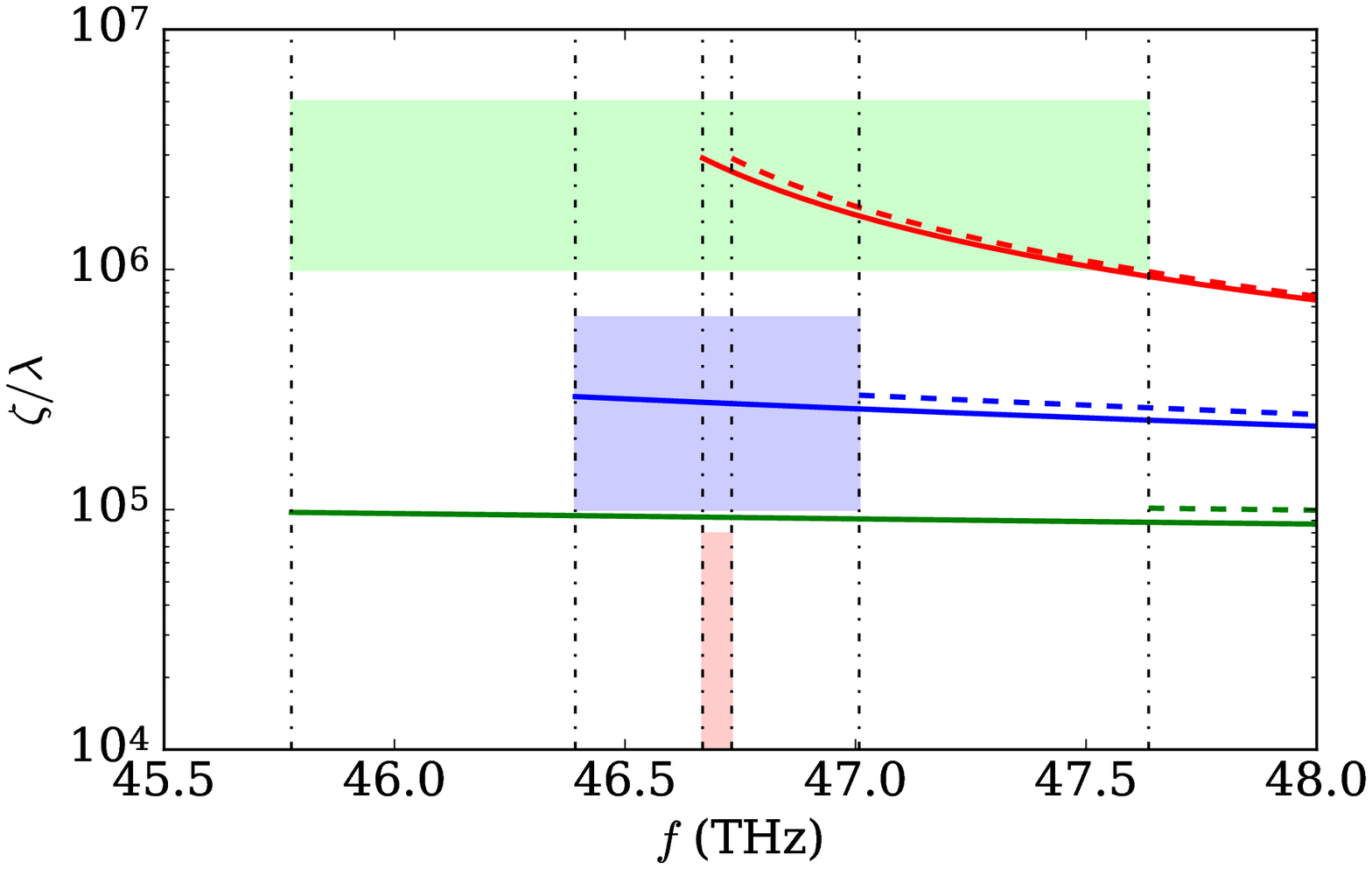}}
\caption{Dispersion curves for the forward (solid curves) and backward (dashed curves) propagating TE surface plasmons along the graphene-ferrite structure shown in Fig.~\ref{fig:graph_ferrite_persp} with anti-parallel magnetic biases. (a) Normalized propagation constants. (b) Normalized propagation length. The ferrites have ferromagnetic resonances $f_0=0.1$~GHz (red), $f_0=1.0$~GHz (blue) and $f_0=3.0$~GHz (green). The corresponding magnetic bias field is $B_0=0.00356$~T (red), $B_0=0.0356$~T (blue), $B_0=0.106$~T (green). A loss factor $\alpha=0.05$ is assumed for the ferrites. For clarity in (a) the blue and green curves are shifted up by 0.1 and 0.2 units, respectively.}
\label{fig:Dispersion}
\end{figure}

Assume the ferrites have anti-parallel magnetic bias $B_0$ as shown in Fig.~\ref{fig:graph_ferrite_persp}, saturation magnetization $M_s$ and loss factor $\alpha$. The permeability tensor for such a ferrite substrate has a resonance at microwave frequencies and its components decrease as $1/f$ at higher frequencies~\cite{lax1962microwave, collin2007foundations}. At infrared and optical frequencies, this magnetic effect becomes vanishingly small. However, due to high sensitivity of the TE mode the resulting nonreciprocity is significant. The normalized confinement factor $\alpha/k_0$ is plotted in Fig.~\ref{fig:alpha_k0} for three different magnetic bias values $B_0=0.0035$~T corresponding to a ferromagnetic resonance $f_0=0.1$~GHz in red, $B_0=0.035$~T corresponding to $f_0=1.0$~GHz in blue and $B_0=0.106$~T corresponding to $f_0=3.0$~GHz in green. Solid curves represent forward, and dashed curves backward propagation. As expected the TE mode is interacting nonreciprocally with the magnetically biased structure. In the highlighted frequency bands the TE mode propagates unidirectionally. This frequency band spans several gigahertz to a few terahertz depending on the strength of the magnetic bias. The corresponding phase constants and propagation lengths are plotted in Fig.~\ref{fig:Dispersion}. The forward and backward plasmons undergo slightly different phases and losses as they propagate along the graphene-ferrite structure. In the highlighted frequency bands the isolation between the forward and backward modes is theoretically infinite. Note that the anti-parallel magnetic bias configuration shown in Fig.~\ref{fig:graph_ferrite_persp} may be produced through a longitudinal DC current. For a $10$~$\mu$m wide graphene strip and a magnetic bias $B_0=0.00035$ corresponding to a ferromagnetic resonance $f_0=0.01$~GHz, the required DC current is $28$~mA. For these parameters the unidirectional propagation bandwidth is 6~GHz.

Note that the operation frequency of the structure can be tuned through the chemical potential of graphene. For higher amounts of doping the operation frequency is increased and for smaller amounts of doping it is lowered towards the microwave frequency band. However as the chemical potential is reduced, fabrication effects such as interaction of graphene with the substrate, which may lead to the modification of the Dirac band structure \cite{giovannetti2007substrate, zhou2007substrate} become important and should be taken into account \cite{werra2016te}. For undoped graphene these interactions greatly modify the characteristics of graphene plasmons \cite{werra2016te}. The energy scale of such modifications in the Dirac cones are normally in the order of 5-50~meV \cite{giovannetti2007substrate, zhou2007substrate}. Depending on the fabrication process of the graphene-ferrite structure and the level of chemical potential, calculation of such effects may be necessary. However, such considerations are beyond the scope of this paper.

\section{Conclusions}
We proposed a graphene ferrite structure that discriminates between the TM and TE surface plasmons of graphene using nonreciprocity, the TE surface plasmon mode, in contrast to its TM counterpart, has a specific nonreciprocal signature, propagating unidirectionally. The TM mode interacts reciprocally. The proposed structure may serve as a platform for the experimental demonstration of the existence of currently still elusive TE plasmonic modes in graphene.

\bibliography{ReferenceList1}

\end{document}